\documentclass[fleqn,12pt,twoside]{article}
\usepackage{espcrc1}

\usepackage{rotating}
\def\1{\mbox{l\hspace{-0.53em}1}}

\title{Excited $[{\bf 70, \ell^+}]$ baryons in large $N_c$ QCD}

\author{N. Matagne\address{University of Li\`ege, Institute of Physics B5, Sart Tilman,
B-4000 Li\`ege 1, Belgium}\thanks{e-mail address: nmatagne@ulg.ac.be} and Fl. Stancu\addressmark\thanks{e-mail address: fstancu@ulg.ac.be}}

\date{\today}

\begin{document}
\maketitle

\begin{abstract}
The masses of the positive parity $[{\bf 70},0^+]$ and $[{\bf 70},2^+]$
non-strange baryons are calculated in 
large $N_c$ QCD by considering the most dominant operators in an 
$1/N_c$ expansion. The approach is based on the introduction
of an excited core, obtained after the last particle (an excited quark) has been
removed. Configuration mixing is neglected, for simplicity.  
Although being a sub-leading $1/N_c$ order,  we find that the spin-spin interaction 
plays a dominant role in describing the data. 
The role of $N_c^ 0$ operators is also pointed out. We show how the contribution
of the linear term in $N_c$, of the spin-spin and of the spin-orbit
terms vary with the excitation energy. 
\end{abstract}

\section{Introduction}

The large $N_c$ QCD approach suggested by 't Hooft \cite{HOOFT} 
and considered in detail by Witten \cite{WITTEN} has become
a powerful tool in baryon spectroscopy. The method is based on
the result that, for $N_f$ flavors, the ground state baryons
display an exact SU($2N_f$) spin-flavor symmetry in the
large $N_c$ limit of QCD \cite{DM93}. 
It has been applied with great success
to the ground state baryons ($N = 0$ band), described
by the symmetric representation $\bf 56$ of SU(6), where $N_f$ = 3
\cite{DM93,Jenk1,DJM94,DJM95,CGO94,JL95,DDJM96}. 
For excited baryons this symmetry is broken. However,
based on the observation that excited states can be approximately classified  
as  SU($2N_f$) multiplets, 
considerable efforts have previously been made to analyze the    
excited states belonging to the $[{\bf 70},1^-]$ multiplet ($N = 1$ band)
in the large $N_c$ limit
\cite{CGKM,Goi97,PY,CCGL,CaCa98,BCCG,SGS,SCHAT,Pirjol:2003ye,GSS03,cohen1},
with obvious success. 

The $N = 2$ band contains the multiplets $[{\bf 56'},0^+], [{\bf 56},2^+],
[{\bf 70},0^+], [{\bf 70},2^+]$ and $[{\bf 20},1^+]$. Among them, the baryons
supposed to belong to  
$[{\bf 56},2^+]$ or $[{\bf 56'},0^+]$ have been considered so far
in Refs. \cite{GSS03} and \cite{CC00}.  
The method of Ref. \cite{GSS03} has recently been extended to
higher excitations  belonging to the $[{\bf 56},4^+]$ multiplet
($N=4$ band) \cite{MS1}.  For simplicity, in these studies configuration
mixing has been neglected.

The symmetric representation  ${\bf 56}$ requires a much simpler treatment 
than the mixed representation ${\bf 70}$.
The main reason is that for the symmetric representation it is not 
necessary to distinguish between excited and core quarks. 
This is possible because of the structure of the wave function.
Let us consider the case $[{\bf 56},2^+]$.  For large $N_c$ this multiplet 
becomes $[[{\bf N_c}],2^+]$ \footnote{This is a partition-type notation. 
It is consistent with the 
label  ${\bf 56}$ of the irrep $[3]$ of SU(6).
The dimension of the symmetric representation of SU($2N_f$)
containing $N_c$ particles and
labelled by the partition $[{N_c}]$  is
$d_{[N_c]} = \frac{(2 N_f + N_c - 1)!}{N_c!~(2N_f - 1)!}$. With $N_f =3$
and $N_c = 3$ one recovers {\bf 56}.}.
Its  intrinsic wave function (center of mass
coordinate removed),
written in a harmonic oscillator single particle basis, takes the form
\begin{equation}\label{SYM2}
|[{\bf N_c}], 2^+\rangle =  
\sqrt{\frac{N_c - 1}{N_c}}|[N_c](0s)^{N_c-1}(0d)\rangle +
\sqrt{\frac{1}{N_c-1}}|[N_c](0s)^{N_c-2}(0p)^2\rangle,
\end{equation} 
which shows that for $N_c \rightarrow \infty$ one can neglect
the second term. This cumbersome term,
involves an excited core after the removal of the last particle, 
while in the first term the entire excitation is carried by
one quark, excited to the $\ell = 2$ shell. The remaining symmetric core
of $N_c - 1$ unexcited quarks is simpler than an excited core.
Now, if we consider the spin-orbit 1-body interaction
\begin{equation}\label{SO}
H_{SO}=w(r)\vec{\ell}\cdot\vec{s}~,
\end{equation}
it is easy to show that
\begin{equation}
\langle\Psi|\ell^i s^i|\Psi\rangle =\left\{
\begin{array}{ll}
\mathcal{O}(N_c^0) &  \mathrm{if}\ \chi \ \mathrm{is}\  \mathrm{mixed}\ 
\mathrm{symmetric}\ \mathrm{(MS)} \\
\mathcal{O}(N_c^{-1}) & \mathrm{if}\ \chi \ \mathrm{is}\ \mathrm{symmetric}\
\mathrm{(S)}   
\end{array}
\right.,
\end{equation}
as proved in Ref. \cite{GOITY04} and, in a different manner, in  
the Appendix below. In this equation $\Psi$ and $\chi$ are the total
and the spin-flavor wave functions of the excited baryon respectively.
(Note that $\chi$ and the spatial wave function of the Appendix 
have the same permutation symmetry, thus the same partition.)  
With this result, the spin-orbit operator takes the simple form 
used in Ref. \cite{GSS03}.

The situation with the multiplets $[{\bf 70},0^+]$ and $[{\bf 70},2^+]$
is different.  This can be illustrated by writing 
the orbital part of the intrinsic wave functions  following the same prescription as above. For 
$\ell = 0$ one obtains
\begin{equation}\label{L0}
|[{\bf N_c-1,1}], 0^+\rangle_{\rho,\lambda} =  
\sqrt{\frac{1}{3}}|[N_c-1,1]_{\rho,\lambda}(0s)^{N_c-1}(1s)\rangle 
+\sqrt{\frac{2}{3}}|[N_c-1,1]_{\rho,\lambda}(0s)^{N_c-2}(0p)^2\rangle.
\end{equation}
In the first term $1s$ is the first single particle
radially excited state with $n=1$, $\ell = 0$ 
($N=2n+\ell$). In the second term     
the two quarks are excited to the $p$-shell to get $N=2$. They are coupled to $\ell = 0$. The lower indices 
$\rho$ and  $\lambda$ are constituent quark model notations. They distinguish 
between states which are antisymmetric and symmetric under permutation
of the first two particles, respectively. By analogy, for $\ell = 2$ one has
\begin{equation}\label{L2}
|[{\bf N_c-1,1}], 2^+\rangle_{\rho,\lambda} =  
\sqrt{\frac{1}{3}}|[N_c-1,1]_{\rho,\lambda}(0s)^{N_c-1}(0d)\rangle
 +\sqrt{\frac{2}{3}}|[N_c-1,1]_{\rho,\lambda}(0s)^{N_c-2}(0p)^2\rangle,
\end{equation}
where the two quarks in the $p$-shell are coupled to  $\ell = 2$.
We have obtained the expressions (\ref{L0}) and (\ref{L2}) by using
the procedure developed in Ref. \cite{MOSHINSKY} based on generalized 
Jacobi coordinates\footnote{For consistency with Ref. \cite{MOSHINSKY}
we use the harmonic oscillator notation $|n \ell \rangle $ for single particle
states everywhere in the text.}.
The $N_c$ = 3 case is demonstrated in Ref. 
\cite{FAIMAN}.

One can see that  
the coefficients of the linear combinations (\ref{L0}) and (\ref{L2})
 are independent of $N_c$
which means that both terms have to be considered in the 
large $N_c$ limit. In  Eqs. (\ref{L0}) and (\ref{L2}) the first term can be 
treated as in the $[{\bf 70},1^-]$ sector, \emph{i.e.} as an excited quark
coupled to a ground state core 
\cite{CGKM,Goi97,PY,CCGL,CaCa98,BCCG,SGS,SCHAT,Pirjol:2003ye,GSS03,cohen1}.
The second term will be treated here as an excited quark coupled to an 
excited core\footnote{One can also consider two excited quarks and 
leave the core in the ground state but this situation is more complicated.}.

To our knowledge this 
is the first attempt to incorporate core excitations in the system and
treat the mass operator accordingly. As there are similarities with the 
$[{\bf 70},1^-]$ multiplet,  we view this study as an extension of Refs.
\cite{CGKM,Goi97,PY,CCGL,CaCa98,BCCG,SGS,SCHAT,Pirjol:2003ye,GSS03,cohen1}. 
In practice we shall combine the techniques of Refs. \cite{CCGL} and \cite{SGS}.  

Before ending this section we should recall that in reality the excited 
states are resonances, so they have a finite width. This is not taken into
account in the present approach, similarly to constituent quark models where 
resonances are treated as bound states. However, it is important to notice, that the bound state picture turned out to describe the baryon phenomenology satisfactorily, thus being a rather realistic picture of baryon resonnances \cite{cohen2}. The decay widths have been
considered, for example, in Refs. \cite{cohen1,cohen2}. The conclusion
was that a general large $N_c$ analysis does not predict narrow widths,
which would vanish in the large $N_c$ limit. Contrary, generic large $N_c$ 
counting gives widths of order $N^0_c$.  According to Ref. \cite{cohen1}
the narrowness of the excited states
is an artifact of simple quarks model assumptions used in the 
calculations, as it is the case here.

\section{The wave function}
We shall separately discuss the two distinct configurations entering 
the wave functions (\ref{L0}) and (\ref{L2}).

\subsection{The configurations $(0s)^{N_c-1}(1s)$ and $(0s)^{N_c-1}(0d)$}

In Eqs. (\ref{L0}) and (\ref{L2}) the first term,
containing the  configurations $(0s)^{N_c-1}(1s)$ and $(0s)^{N_c-1}(0d)$
respectively, is similar in structure with the
wave functions of the $[{\bf 70},1^-]$ multiplet ($N_c = 3$), the difference  
being that the quark is now excited to the $N = 2n + \ell =2 $ band ($\ell = 0$
or $\ell = 2$) instead of $N=\ell = 1$.

The configurations $(0s)^{N_c-1}(1s)$ and $(0s)^{N_c-1}(0d)$
are then described by a wave function written in the notation of Ref. \cite{CCGL} as
\begin{eqnarray}\label{CORE}
\lefteqn{|JJ_3;II_3;(\ell=\ell_q;S=I+\rho)\rangle  = } \nonumber \\ 
\sum_{m_\ell,m_S,m_1,m_2,\alpha_1,\alpha_2}  
&  \left(\begin{array}{cc|c}
	\ell_q    &    S   & J   \\
	m_{\ell_q}  &   m_S  & J_3 
      \end{array}\right)
\left(\begin{array}{cc|c}
	S_c    &    \frac{1}{2}   & S   \\
	m_1  &         m_2        & m_S 
      \end{array}\right)
\left(\begin{array}{cc|c}
	I_c    &  \frac{1}{2}  & I   \\
	\alpha_1 &  \alpha_2    & I_3 
      \end{array}\right) \nonumber \\
  \sum_{\eta=\pm  1} & c_{\rho,\eta} |S_c=I_c=I+\frac{\eta}{2};m_1,\alpha_1\rangle \otimes|\frac{1}{2};m_2,\alpha_2\rangle\otimes 
|\ell_q,m_{\ell_q}\rangle,      
\end{eqnarray}
where $S$ and $I$ are the total spin and isospin of the baryon,
$\ell = \ell_{q}$ is the angular momentum of the excited quark, 
$S_c$ and $I_c$ are the core spin and isospin respectively, 
$\rho \equiv S-I = \pm 1,0$ and $\eta/2 \equiv I_c-I=\pm 1/2$.  
The coefficients $c_{\rho,\eta}$ will be presented below. The coupling
coefficients are SU(2) coefficients, as we consider here only 
non-strange baryons ($N_f = 2$).

\subsection{The configurations $(0s)^{N_c-2}(0p)^2$ }

In order to better understand the  $1/N_c$ counting it is useful to
rewrite the orbital part of the wave function using the fractional
parentage technique \cite{book} by which the last particle is decoupled. 
Here we consider the only case treated in the literature,
where the mixed symmetric state belonging to the representation
$[N_c-1,1]$ has the last particle in the second row of the corresponding
Young diagram. 
Then, the second term of (\ref{L0}) or (\ref{L2}) takes the form
\begin{eqnarray}
|[N_c-1,1]_{\rho,\lambda} (0s)^{N_c-2}(0p)^2,\ell^+ \rangle  &=& 
\sqrt{\frac{N_c-2}{N_c}} \Psi_{[N_c-1]}\left((0s)^{N_c-2}~(0p)\right) \phi_{[1]}(0p) \nonumber \\
& & -\sqrt{\frac{2}{N_c}} \Psi_{[N_c-1]}\left((0s)^{N_c-3}~(0p)^2\right)\phi_{[1]}(0s)~,
\end{eqnarray}
the decoupling being valid both for $\ell = 0$ and 2. Here all states are normalized. 
The first factor in each term in the right-hand side is a symmetric $(N_c-1)$-particle wave function
and $\phi_{[1]}$ is a one particle wave function associated to
the $N_c$-th particle. One can see that 
for large $N_c$ the  coefficient of the first term is $\mathcal{O}(1)$ and of the second $\mathcal{O}(N_c^{-1/2})$.
Then, in the large  $N_c$ limit, one can  neglect the second term and take into account
only  the first term in the wave function, 
where the $N_c$-th particle has an $\ell = 1$ excitation.
A similar decomposition can be made for a symmetric state $[N_c]$ or
a mixed symmetric state $[N_c-1,1]$
containing one excited quark only (see Appendix). In that case
one can immediately recover the considerations developed in 
Ref. \cite{GOITY04} for the spin-orbit or other operators.

When both the decoupled quark and the core are excited,
the wave function takes the form
\begin{eqnarray}\label{EXCORE}
\lefteqn{|JJ_3;II_3;(\ell;S=I+\rho)\rangle  =}\nonumber \\
&& \sum_{m_q,m_c,m_\ell,m_S,m_1,m_2,\alpha_1,\alpha_2}  
  \left(\begin{array}{cc|c}
	\ell    &    S   & J   \\
	m_\ell  &   m_S  & J_3 
      \end{array}\right)
\left(\begin{array}{cc|c}
	\ell_q    &  \ell_c   & \ell   \\
	m_q  &    m_c    & m_\ell 
      \end{array}\right) \nonumber \\  
&& \left(\begin{array}{cc|c}
	S_c    &    \frac{1}{2}   & S   \\
	m_1  &         m_2        & m_S 
      \end{array}\right)
\left(\begin{array}{cc|c}
	I_c    &  \frac{1}{2}  & I   \\
	\alpha_1 &  \alpha_2    & I_3 
      \end{array}\right) \nonumber \\
&&\sum_{\eta=\pm  1}   c_{\rho,\eta} |S_c=I_c=I+\frac{\eta}{2};m_1,\alpha_1\rangle \otimes|\frac{1}{2};m_2,\alpha_2\rangle 
\otimes|\ell_q,m_q\rangle  \otimes |\ell_c,m_c\rangle,     
\end{eqnarray}
which is a generalization of (\ref{CORE}) still for non-strange baryons.  
It contains an extra SU(2) Clebsch-Gordan coefficient which couples the angular momentum $\ell_q$ of the excited quark to the angular momentum $\ell_c$ of the excited core.
The coefficients $c_{\rho,\eta}$ are the same in (\ref{CORE}) and (\ref{EXCORE}). They are defined in Refs. \cite{CCGL} and \cite{SGS}. We recall that for the symmetric (SYM) $[N_c]$ and 
the mixed (MS) $[N_c-1,1]$ representations of the permutation group $S_{N_c}$,
the non-vanishing coefficients $c_{\rho,\eta}$ 
are given by 
\begin{equation}\label{coeff1}
c^{\mathrm{MS}}_{\pm,\pm} = 1,
\end{equation} 
\begin{equation}\label{coeff2}
c^{\mathrm{MS}}_{0+} = \sqrt{\frac{S [N_c + 2(S + 1)]}{N_c (2 S + 1)}}, \\
~~~~~c^{\mathrm{MS}}_{0-} = - \sqrt{\frac{(S + 1)[N_c - 2 S]}{N_c (2 S + 1)}}, 
\end{equation}
\begin{equation}\label{coeff3}
c^{\mathrm{SYM}}_{0+} = - c^{\mathrm{MS}}_{0-}, ~~~~~ c^{\mathrm{SYM}}_{0-} = c^{\mathrm{MS}}_{0+}.
\end{equation}
In Ref. \cite{CCGL} they were defined as elements of an orthogonal basis
rotation. In terms of group theory language these coefficients can be 
identified with isoscalar
factors of the permutation group \cite{book,ISOSC}. They are factors of the
Clebsch-Gordan coefficients needed in the flavor-space inner product
and are related to the position of the last particle in the 
corresponding Young tableaux.

\section{The Mass Operator}

For the $[{\bf 70},\ell^+]$ sector the building blocks which form the operators
entering in the mass formula consist of the excited core operators $\ell^i_c$, 
$S^i_c$, $T^a_c$ and $G^{ia}_c$ and the excited quark operators $\ell^i_q$, 
$s^i$, $t^a$ and $g^{ia}$. We also introduce the tensor 
operator\footnote{The irreducible spherical tensors are defined according to  
Ref. \cite{BRINK}.}
\begin{equation}\label{TENSOR}
\ell^{(2)ij}_{ab}=\frac{1}{2}\left\{\ell^i_a,\ell^j_b\right\}
-\frac{1}{3}\delta_{i,-j}\vec{\ell}_a\cdot\vec{\ell}_b~,
\end{equation}
with $a=c$, $b=q$ or vice versa or $a=b=c$ or $a=b=q$. For simplicity 
when $a=b$, we shall use a single index $c$, for the core, and $q$ for the 
excited quark so that the operators are $\ell^{(2)ij}_c$ and $\ell^{(2)ij}_q$ 
respectively. The latter case represents the tensor 
operator used in the analysis of the $[{\bf 70},1^-]$ multiplet (see \emph{e.g.} 
Ref. \cite{CCGL}). 

\begin{table}[htb]
\caption{List of operators and the coefficients resulting from the fit with $\chi^2_{\rm dof}  \simeq 0.83$.}
\label{operators}
\renewcommand{\arraystretch}{1.2} 
\begin{tabular}{llrrl}
\hline
\hline
Operator & \multicolumn{4}{c}{Fitted coef. (MeV)}\\
\hline
\hline
$O_1 = N_c \ \1 $                                   & \ \ \ $c_1 =  $  & 555 & $\pm$ & 11       \\
$O_2 = \ell_q^i s^i$                                & \ \ \ $c_2 =  $  &   47 & $\pm$ & 100    \\
$O_3 = \frac{3}{N_c}\ell^{(2)ij}_{q}g^{ia}G_c^{ja}$ & \ \ \ $c_3 =  $   & -191 & $\pm$ & 132  \\
$O_4 = \frac{1}{N_c}(S_c^iS_c^i+s^iS_c^i)$          & \ \ \ $c_4 =  $  &  261 & $\pm$ &  47  \\
\hline \hline
\end{tabular}
\end{table}


\begin{table}[htb]
\caption{Matrix elements of $N$.}
\label{NUCLEON}
\renewcommand{\arraystretch}{1.2}
\begin{tabular}{lcccc}
\hline 
\hline
   &  \hspace{ .3 cm} $O_1$ \hspace{ .3 cm}  & \hspace{ .3 cm} $O_2$  \hspace{ .3 cm} & \hspace{ .3 cm} $O_3$  \hspace{ .3 cm}&  \hspace{ .3 cm} $O_4$  \hspace{ .3 cm}  \\
\hline
$^4N[{\bf 70},2^+]\frac{7}{2}^+$  &  $N_c$   &  $\frac{2}{3}$    & $-\frac{1}{6N_c}(N_c+1)$ & $\frac{5}{2N_c}$ \\
$^2N[{\bf 70},2^+]\frac{5}{2}^+$  &  $N_c$   &  $\frac{2}{9N_c}(2N_c-3)$ & 0  &   $\frac{1}{4N_c^2}(N_c+3)$  \\
$^4N[{\bf 70},2^+]\frac{5}{2}^+$  &  $N_c$   &  $-\frac{1}{9}$  & $\frac{5}{12N_c}(N_c+1)$  &   $\frac{5}{2N_c}$ \\
$^4N[{\bf 70},0^+]\frac{3}{2}^+$  &  $N_c$   &  0    & 0 &   $\frac{5}{2N_c}$ \\
$^2N[{\bf 70},2^+]\frac{3}{2}^+$  &  $N_c$   &   $-\frac{1}{3N_c}(2N_c-3)$   & 0 &  $\frac{1}{4N_c^2}(N_c+3)$ \\
$^4N[{\bf 70},2^+]\frac{3}{2}^+$  &  $N_c$   &   $-\frac{2}{3}$   & 0 &   $\frac{5}{2N_c}$ \\
$^2N[{\bf 70},0^+]\frac{1}{2}^+$  &  $N_c$   &   0   & 0 &  $\frac{1}{4N_c^2}(N_c+3)$ \\
$^4N[{\bf 70},2^+]\frac{1}{2}^+$  &  $N_c$   &    $-1$   &  $-\frac{7}{12N_c}(N_c+1)$ &  $\frac{5}{2N_c}$ \\
\hline
\hline
\end{tabular}
\end{table}



\begin{table}[htb]
\caption{Matrix elements of $\Delta$.}
\label{DELTA}
\renewcommand{\arraystretch}{1.2}
\begin{tabular}{lcccc}
\hline 
\hline
   &  \hspace{ .3 cm} $O_1$ \hspace{ .3 cm}  & \hspace{ .3 cm} $O_2$  \hspace{ .3 cm} &  \hspace{ .3 cm} $O_3$  \hspace{ .3 cm} &  \hspace{ .3 cm} $O_4$  \hspace{ .3 cm}  \\
\hline

$^2\Delta[{\bf 70},2^+]\frac{5}{2}^+$  &   $N_c$   &  $-\frac{2}{9}$ & 0  & $\frac{1}{N_c}$ \\
$^2\Delta[{\bf 70},2^+]\frac{3}{2}^+$  &   $N_c$   &  $\frac{1}{3}$  & 0  & $\frac{1}{N_c}$ \\
$^2\Delta[{\bf 70},0^+]\frac{1}{2}^+$  &   $N_c$   &  0 & 0 & $\frac{1}{N_c}$ \\

\hline
\hline
\end{tabular}
\end{table}


We apply the $1/N_c$ 
counting rules presented in  Refs. \cite{CCGL} and \cite{SGS} and use their 
conclusions in selecting the most dominant operators in the practical
analysis. For non-strange baryons, 
Table I of Ref. \cite{CCGL} gives a list of 18 linearly independent operators.
If the core is excited the number of operators appearing in the mass formula
is much larger. 
However 
due to lack of data, here we have to consider a restricted list. 
The selection is suggested by the conclusion  
of Ref. \cite{CCGL}, $(N_f=2)$  and of Ref. \cite{SGS} 
$(N_f = 3)$, that only 
 a few operators, of some specific structure, bring
a dominant
contribution to the mass. Following the notation of Ref. \cite{SGS} these are
$O_1, O_2, O_3$ and $O_4$ exhibited here in Table I. The first is the trivial
operator of order $\mathcal{O}(N_c)$. The second is the 1-body part of the spin-orbit 
operator of order $\mathcal{O}(1)$ which acts on the excited quark.
The third is a composite 2-body operator formally of order $\mathcal{O}(1)$ as well. 
It involves 
the tensor operator (\ref{TENSOR}) acting on the excited quark and the SU(6)
generators $g^{ia}$ acting on the excited quark and $G^{ja}_c$ acting on the
core. The latter is a coherent operator which introduces an extra power $N_c$
so that the order of $O_3$ is $\mathcal{O}(1)$, as it can be seen from Table  
\ref{NUCLEON}.
In order to take into 
account its contribution we have applied the rescaling introduced 
in Ref. \cite{SGS} which consists in introducing a multiplicative factor of 3. 
Without this factor the coefficient $c_3$ becomes too large,
as noticed in Ref. \cite{SGS} \footnote{
Alternatively the factor 3 could be included in the definition (\ref{TENSOR})
of the tensor operator, as sometimes done in the literature. In practice
what it matters is the product $c_i O_i$.}. 
The dynamics of the operator $O_3$ is less
understood. Previous studies \cite{CCGL,SGS}
speculate about its connection to a flavor exchange
mechanism \cite{GLOZMAN,GEORGI} related to long distance meson exchange 
interactions. Finally, the last operator is the spin-spin interaction, the only one of
 order $\mathcal{O}(1/N_c)$ which we consider here. Higher order operators are neglected.
Accordingly the mass operator of the $[{\bf 70},\ell^+]$ multiplet
is approximated by
\begin{equation}
M_{[{\bf 70},\ell^+]} = \sum_{i=1}^4 c_i O_i~,
\end{equation}
where the coefficients $c_i$ have to be found in a numerical fit
to the available data, as described below.
The diagonal matrix elements of the operators $O_i$ $(i = 1,...,4)$ are given in
Tables  \ref{NUCLEON} and \ref{DELTA}. They have been obtained 
from the wave functions (\ref{L0}) and (\ref{L2}).
Each matrix element contains the additional contribution  
of the two terms of the wave function. We derived general analytic formulae for each case in
terms of $N_c$.  
For the first term which has a ground state core,
we used Eq. (\ref{CORE})  to recover the matrix elements of Appendix A 
of Ref. \cite{CCGL}.
For the second term, using Eq. (\ref{EXCORE}), we have derived analytic 
expressions
which generalize those of  Ref. \cite{CCGL} to an excited core
with angular momentum $\ell_c \neq 0$. Taking $\ell_c = 0$ we reobtained 
the corresponding expressions of Ref. \cite{CCGL}. Details will be 
given elsewhere.

\section{Fit and conclusions}

\begin{sidewaystable}
\caption{The partial contribution and the total mass (MeV) predicted by the $1/N_c$ expansion as compared with the empirically known masses.}
\label{multiplet}
\renewcommand{\arraystretch}{1.2}
\begin{tabular}{lrrrrccl}\hline \hline
                    &      \multicolumn{5}{c}{$1/N_c$ expansion results}        &    &                     \\ 
\cline{2-6}		    
                    &      \multicolumn{4}{c}{Partial contribution (MeV)} & \hspace{.7cm} Total (MeV)  \hspace{.5cm}  & \hspace{.3cm}  Empirical (MeV) \hspace{2cm}&   Name, status \hspace{.0cm} \\
\cline{2-5}
                    &   \hspace{.3cm}   $c_1O_1$  & \hspace{.3cm}  $c_2O_2$ & \hspace{.4cm}$c_3O_3$ &\hspace{.3cm}  $c_4O_4$   &    &        \\
\hline
$^4N[{\bf 70},2^+]\frac{7}{2}^+$        & 1665 & 31 & 42 & 217 &      $ 1956\pm95$  & $2016\pm104$ &  $F_{17}(1990)$**  \\
$^2N[{\bf 70},2^+]\frac{5}{2}^+$      & 1665 & 10   & 0 & 43 &      $1719\pm34 $  &    \\

$^4N[{\bf 70},2^+]\frac{5}{2}^+$   & 1665 & -5  & -106 &  217 &     $ 1771\pm88$  & $1981\pm200$ & $F_{15}(2000)$**  \\
$^4N[{\bf 70},0^+]\frac{3}{2}^+$   & 1665  & 0   & 0 & 217 &    $1883\pm17$  & $1879\pm17$ &  $P_{13}(1900)$** \\
$^2N[{\bf 70},2^+]\frac{3}{2}^+$  &   1665   &   -16  & 0 &  43   & $1693\pm42$  &                                \\
$^4N[{\bf 70},2^+]\frac{3}{2}^+$     &  1665    &   -31  & 0 & 217    & $1851\pm69$  &                                   \\
$^2N[{\bf 70},0^+]\frac{1}{2}^+$   &   1665   &   0  &  0 & 43  &   $1709\pm25$  &     $1710\pm30$          &    $P_{11}(1710)$***                 \\
$^4N[{\bf 70},2^+]\frac{1}{2}^+$   & 1665  & -47   & 149 & 217 &     $1985\pm26 $  &   $1986\pm26$ &  $P_{11}(2100)$*\\
\hline
$^2\Delta[{\bf 70},2^+]\frac{5}{2}^+$  &  1665    &   -10  &  0 & 87  &    $1742\pm29$  &  $1976\pm237$             &  $P_{35}(2000)$**                 \\
$^2\Delta[{\bf 70},2^+]\frac{3}{2}^+$     &   1665   &  16   &  0 & 87    &  $1768\pm38$  &                                   \\
$^2\Delta[{\bf 70},0^+]\frac{1}{2}^+$   &   1665   & 0    &  0 &  87 &   $1752\pm19$  &   $1744\pm36$            &   $P_{31}(1750)$*                  \\
\hline
\hline
\end{tabular}
\end{sidewaystable}


In Table \ref{multiplet} we present the masses of the resonances which we have
interpreted as belonging to the $[{\bf 70},0^+]$ or to the $[{\bf 70},2^+]$ multiplet.
For simplicity, mixing of multiplets is neglected in this first attempt.
The resonances shown in column 8, correspond to
either  three stars (``very likely") or to two stars (``fair") or to
one star (``poor") status,
according to Particle Data Group (PDG) \cite{EIDELMAN}.
Therefore we used the full listings  to determine a  
mass average in each case. The experimental error to the mass was calculated 
as the quadrature of two uncorrelated errors, one being the average 
error from the same references and the other was the difference
between the average mass and the farthest observed mass.  
For the $P_{11}(2100)$* resonance we report results from
fitting the experimental value
of Ref. \cite{THOMA}, as being more recent than the average over the
PDG values. Note that the observed mass of Ref. \cite{THOMA} is in
agreement with the recent coupled channel analysis of Manley and
Saleski \cite{MANLEY}.

Several remarks are in order. Due to its large error in the mass,
the resonance $F_{15}$(2000) could be either described by the
$|^2N[{\bf 70},2^+]5/2^+\rangle$ state or by  the $|^4N[{\bf 70},2^+]5/2^+\rangle$ state 
(inasmuch as they appear
separated by about 60-70 MeV only, in quark model studies, see, \emph{e.g.}, \cite{IK,SS}).
Here we identified $F_{15}$(2000) with the $|^4N[{\bf 70},2^+]5/2^+\rangle$ state
because it gives a better fit. Regarding the $F_{35}$(1905) resonance
there is also some ambiguity. In Ref. \cite{GSS03} it was identified 
as a $|^4\Delta[{\bf 56},2^+]5/2^+\rangle$ state following Ref. \cite{IK}, but in
Ref. \cite{MS1} the interpretation $|^4\Delta[{\bf 70},2^+]5/2^+\rangle$ was preferred 
due to a better $\chi^2$ fit and other considerations related to the decay 
width. Here we return to the identification made in Ref. \cite{GSS03}
and assign the $|^4\Delta[{\bf 70},2^+]5/2^+\rangle$ state to the second 
resonance from this sector, namely $F_{35}$(2000), as indicated in Table \ref{multiplet}.
We hope that an analysis based on configuration
mixing and improved data could better clarify the resonance assignment
in this sector in the future. Presently the resulting 
$\chi^2_{\mathrm{dof}}$ is about 0.83 and the fitted values of $c_i$
are given in Table \ref{operators}.
Besides the seven fitted masses Table \ref{multiplet} also contains few 
predictions.

We found that the contributions of
$S^i_c S^i_c$ and $s^i S^i_c$ are nearly equal when treated as
independent operators. Therefore, for simplicity, we assumed that they
have the same coefficient in the mass operator. 
One can see that the spin-spin interaction given by $O_4$ is the  dominant  
interaction, as in the $[{\bf 56},2^+]$ multiplet \cite{GSS03} or 
in the $[{\bf 56},4^+]$ multiplet \cite{MS1}. 
Thus the main contributions to the mass come from  $O_1$ and $O_4$.
It is remarkable that $c_1$ and $c_4$ of the multiplets 
$[{\bf 56},2^+]$ and $[{\bf 70},\ell^+]$, both located in the $N = 2$ band, are very close to each other. 
In terms of the present notation the result of Ref. \cite{GSS03} for $[{\bf 56},2^+]$ is 
$c_1 = 541 \pm 4$ MeV and $c_4 = 241 \pm 14$ MeV as compared to $c_1 = 555 \pm 11$ MeV and 
$c_4 = 261 \pm 47$ MeV here. Such similarity  gives
confidence in the large $N_c$ approach and in the present fit. 
\begin{figure}
\begin{center}
\includegraphics[width=10cm]{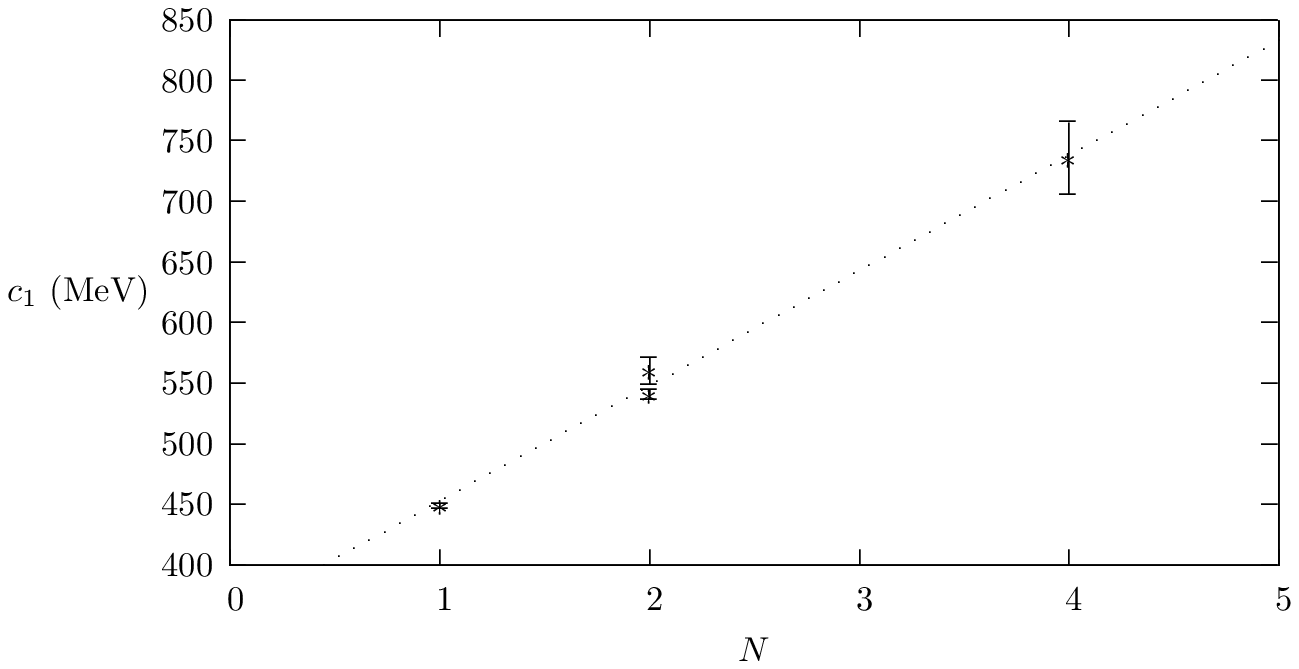} \\
\vspace{0.5cm}
\includegraphics[width=10cm]{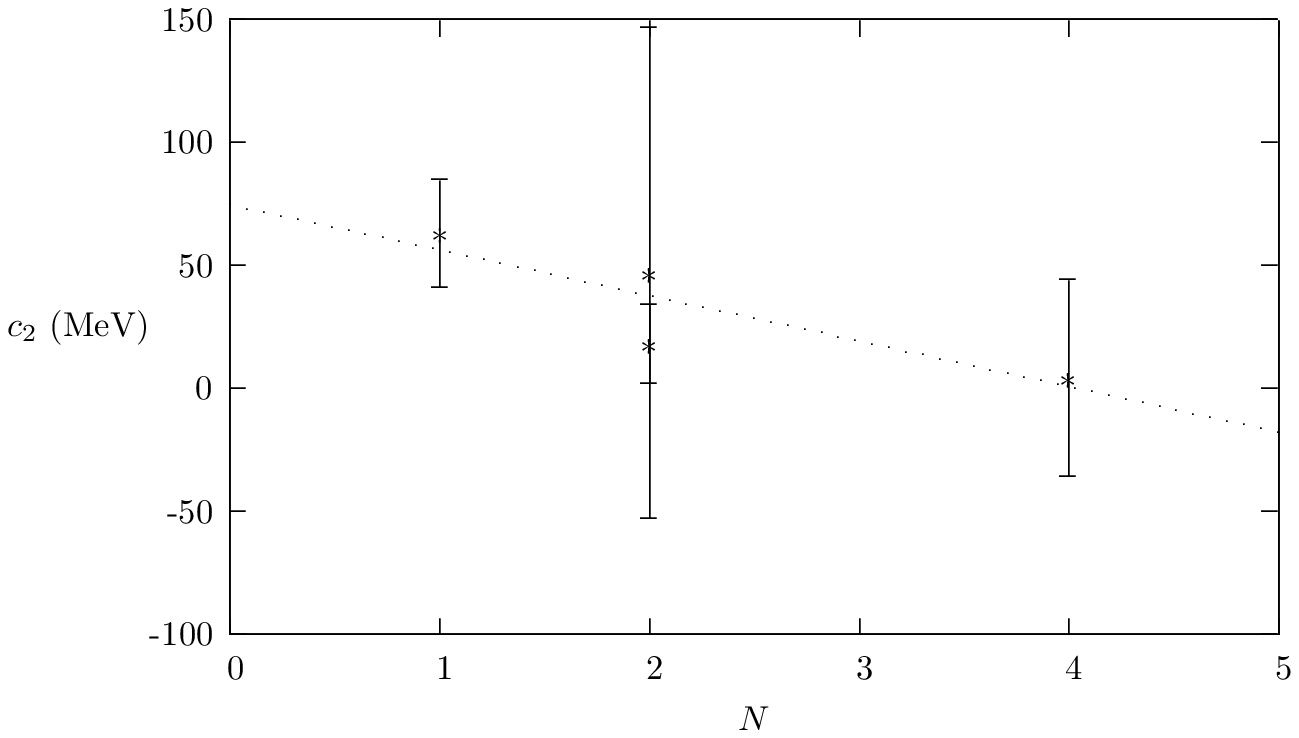} \\
\vspace{0.5cm}
\includegraphics[width=10cm]{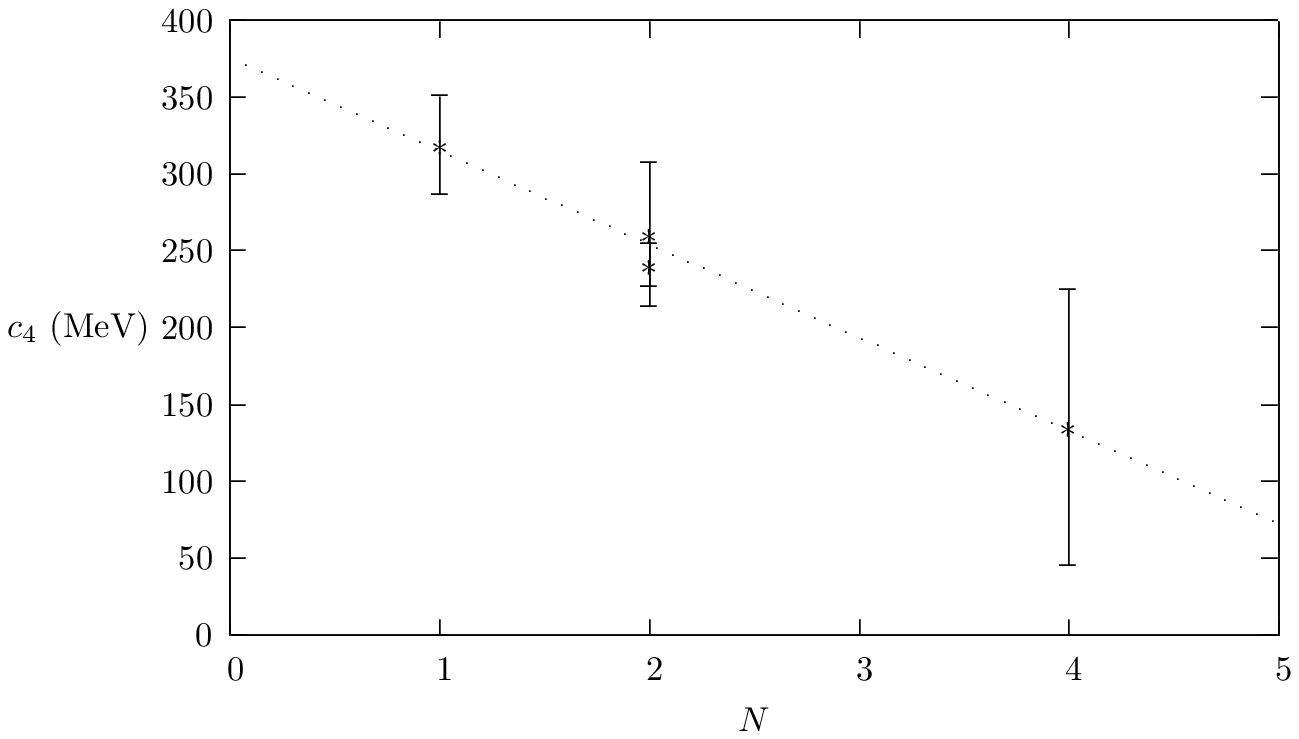}
\vspace{-1cm}
\caption{The coefficients $c_i$ vs $N$ from various sources:
for $N = 1$ from Ref. \cite{SGS}
for $N = 2$ from Ref. \cite{GSS03} (lower values) and present work (upper values),
for $N = 4$ from Ref. \cite{MS1}. The straight lines are to guide the eye.}
\end{center}
\end{figure}
In Fig. 1 the presently known values of $c_1, c_2$ and $c_4$ with error bars 
are represented for the excited bands studied within the large $N_c$
expansion: $N = 1$ is from Ref. \cite{SGS},  $N = 2$ from Ref. \cite{GSS03} and
from the present work, $N = 4$ from Ref. \cite{MS1}. Extrapolation to higher 
energies, $N > 4$, suggests that the contribution of the spin dependent operators
would vanish, while the linear term in $N_c$,
which in a quark model picture would contain the free mass term, the kinetic 
and the confinement energy, would carry the entire
excitation.  Such a behaviour 
gives a deeper insight into the large $N_c$ mass operator
and is consistent with
the intuitive picture developed in Ref. \cite{LYG} where at high energies the 
spin dependent interactions vanish as a consequence of the 
chiral symmetry restoration.
   
Finally note that the contributions of  $O_2$ 
and $O_3$ lead to large errors in the coefficients $c_i$ obtained in the $\chi^2$ fit,
which could possibly be removed with better data.
The operator $O_3$ containing the tensor term plays an important
role in the reduction of $\chi^2_{\mathrm{dof}}$ and it should be further investigated. 

In conclusion we believe that the mass fit of the $[{\bf 70},\ell^+]$ baryons is
encouraging in the present form. 
It is consistent with the conclusions of previous studies that the dominant 
interaction is of a spin-spin nature. It would be interesting to 
extend this analysis such as to incorporate configuration mixing with the 
$[{\bf 56},2^+]$ states. In this way one could obtain a complete description of 
the $N=2$ band baryons, leaving alone the 
$[{\bf 20},1^+]$ multiplet for which no candidate has been found so far. 
On the other hand improved experimental data is highly desirable.

\vspace{1cm}
\centerline{\bf Acknowledgments}

We are most grateful to
Aneesh Manohar and Carlos Schat for a careful reading of the manuscript and good advise
for the numerical fit. Useful correspondence with 
Jos\'e Goity  and Norberto Scoccola is gratefully
acknowledged. We thank Pierre Stassart for useful comments. 
Both of us benefited of hospitality at ECT* Trento where we could have fruitful discussions on this subject.
The work of one of us (N. M.) was supported by the Institut Interuniversitaire 
des Sciences Nucl\'eaires (Belgium).

\vspace{1cm}
\centerline{\bf Appendix}

Here we express a wave function of a given symmetry described by
the partition $[f]$ by using fractional parentage coefficients.
They are related to the isoscalar factors of the permutation
group \cite{book}. For one-body operators we need one-body 
fractional parentage coefficients. In this way one can 
decouple the last particle from the rest. 
In the simple case where the spatial wave function contains only one
excited quark, for example having the structure $(0s)^{N_c-1}(0d)$ 
(two units of orbital excitation), and symmetry $[N_c-1,1]$
one can show that 
\begin{eqnarray}
|[N_c-1,1] (0s)^{N_c-1}(0d),2^+ \rangle  &=& 
\sqrt{\frac{N_c-1}{N_c}} \Psi_{[N_c-1]}(0s)^{N_c-1}\phi_{[1]}(0d) \nonumber \\
& & -\sqrt{\frac{1}{N_c}} \Psi_{[N_c-1]}\left((0s)^{N_c-2}(0d)\right)\phi_{[1]}(0s)~.
\end{eqnarray} 
In the case of the spin-orbit operator one can see that only the first term
contributes (the operator acts on the last particle only). Its matrix
element is proportional to the square 
of the coefficient of the first term, \emph{i.e.} with  ${\frac{N_c-1}{N_c}}$  
which for large $N_c$ gives to the spin-orbit the order $\mathcal{O}(1)$ in
powers of $1/N_c$.  

On the other hand for a symmetric state $[N_c]$
with the same structure one obtains
\begin{eqnarray}
|[N_c] (0s)^{N_c-1}(0d),2^+ \rangle  &=& 
\sqrt{\frac{1}{N_c}} \Psi_{[N_c-1]}(0s)^{N_c-1}\phi_{[1]}(0d) \nonumber \\
& & + \sqrt{\frac{N_c-1}{N_c}} \Psi_{[N_c-1]}\left((0s)^{N_c-2}(0d)\right)\phi_{[1]}(0s)~,
\end{eqnarray} 
which implies that the spin-orbit operator is of order $\mathcal{O}(1/N_c)$.
Both results are in agreement with Eq. (\ref{SO}).


\end{document}